\documentclass[prl,twocolumn,showpacs,amsmath,amssymb]{revtex4}
\usepackage{bm}
\usepackage{graphicx}
\begin{document}
\title{Correlating supercooling limit and glass-like arrest of kinetics for disorder-broadened 1st order transitions: relevance to phase separation.}
\author{P. Chaddah and A. Banerjee} 
\affiliation{UGC-DAE Consortium for Scientific Research (CSR)\\ University Campus, Khandwa Road, Indore 452 017, India.}
\author{S. B. Roy}
\affiliation{Magnetic and Superconducting Materials Section, Raja Ramanna Centre for Advanced Technology, Indore 452013, India.} 
\date{\today}
\begin{abstract}
Coexisting ferromagnetic (FM) and antiferromagnetic (AFM) phases over a range of temperature (T) and magnetic field (H), have been reported in many materials. The 1st order FM-AFM transition is completed over a broad T (or H) range; this is ascribed to a landscape of free-energy density in systems with frozen disorder.  Kinetic arrest, akin to glass transition, has been reported in doped CeFe$_2$, La-Pr-Ca-Mn-O, Gd$_5$Ge$_4$, etc., such that the coexisting fraction is frozen.  The de-arrest, or melting, of this glass-like arrested state is seen to occur over a range of temperatures, implying a landscape (of T$_K$ or T$_g$).  We argue that measuring magnetization along various specific paths in H-T space can help infer whether the landscapes of free-energy and for kinetic arrest are correlated. This will help to determine whether disorder affects glass formation, and the underlying 1st order transition, in contrasting (or similar) ways.        
\end{abstract}

\maketitle
First order phase transitions (FOPT) that can be caused by either varying temperature or by varying H are of current interest because these occur in various  magnetocaloric materials, colossal magnetoresistance materials, magnetic shape-memory alloys etc., and are also believed to be the cause for the functional properties of these materials \cite{dag}. In the absence of disorder, the FOPT occurs at a sharply defined (H${_C}$, T${_C}$) line in the 2-control variable (H, T) space. Many of these functional magnetic materials are multi-component systems whose properties become more interesting under substitution. Such substitutions are an intrinsic source of frozen disorder. Early theoretical arguments of Imry and Wortis \cite{imry} showed that such samples would show a disorder-broadened transition, with a spatial distribution of the (H${_C}$, T${_C}$) line across the sample. The first visual realization of such a local variation was provided by Soibel et al. \cite{soibel} for the vortex melting transition. A similar visual realization for an antiferromagnetic (AFM) to ferromagnetic (FM) transition, in Ru-doped CeFe$_2$, was provided by Roy et al. \cite{roy} for the FOPT being caused by variation of temperature (with field held constant), and also by the variation of field (with temperature held constant). The occurrence of a landscape of free energy densities, and a spread of local (H${_C}$, T${_C}$) values across the samples, would result in the (H${_C}$, T${_C}$) line being broadened into a band for samples with frozen disorder. The spinodal lines corresponding to the limit of supercooling (H*, T*) and corresponding to the limit of superheating (H**, T**), would also be broadened into bands \cite{manekar}. Each of these bands corresponds to a quasi-continuum of lines; each line corresponds to a region of the disordered sample with length-scale of the order of the correlation length. Since the correlation length is finite, the number N of these regions (and thus of lines in the band) could be large but would be finite.  We propose to focus on the quasi-continuum of lines forming the band.

In a parallel development the disorder-broadened FOPTs and their slow dynamics or hindered kinetics has been under investigation \cite{manekar,singh,chatto1,chatto2,sharma,ghive}. It is believed that in various materials the kinetics is actually arrested (on experimental or laboratory time scales) and a `glass' is formed \cite{ref1} , and specifically, the FOPT is fully or partially, arrested at low temperatures. The arrest occurs as one cools, and this state melts or gets `de-arrested' over a range of temperature \cite{chatto2,sharma,ghive} as one warms.  In addition to CeFe${_2}$, these effects have been seen in Gd${_5}$Ge${_4}$ \cite{chatto1}, La-Pr-Ca-Mn-O \cite{sharma,ghive}, Nd-Sr-Mn-O \cite{kuwa,tokura1}, Nd$_7$Rh$_3$ \cite{sampath} (see also references [5 and 6]). If such a kinetic arrest were to occur below a (H${_K}$, T${_K}$) line in the pure system, the disordered system would have a (H${_K}$, T${_K}$) band formed out of the quasi-continuum of (H${_K}$, T${_K}$) lines. Each line would correspond to a local region of the sample. If the lines in each of these bands correspond to different regions of the sample, can one seek a correlation between the position of a line in the (H${_K}$, T${_K}$) band and the position of the corresponding (i.e. from the same region) line in the (H*, T*) band?  We now consider isothermal M vs H measurements, after field-cooling (FC) in various fields, as a means of answering this question. 
 
Let the high-T state be AFM and the low-T ground state be FM. Since a higher field favours the FM phase, the (H*, T*)  band moves to higher T at higher H. We take the (H${_K}$, T${_K}$) band to lie above the (H*, T*) band at zero field (to be consistent with the observation in La-Pr-Ca-Mn-O that zero-field cooling (ZFC) results in fully arrested AFM), and to rise to higher H as T is lowered (since FC in large field gives FM). This is shown in the schematic in Fig. 1(a). The bands are actually a quasi-continuum of N-lines, N=4 having been depicted in the schematic.  We thus have 3 groups (say x, y and z) of these N lines in each of the T* and T$_K$ bands.  We assume that the high-T end of the (H*, T*) band and the low-T end of the (H${_K}$, T${_K}$) band correspond to the same local region (i.e. the two bands are anti-correlated), as indicated in the schematic. In path 1 we cool in zero-field, all T* lines lie below T$_K$ lines, and entire sample is frozen in AFM at T$_O$. As H is raised at T$_O$, regions in group x will start de-arresting at the field marked by horizontal arrow 1. As H rises, group y and then group z regions will de-arrest into FM phase. For FC along path 2, the regions  in group x will transform from AFM to FM, but the regions in group y and z are arrested in AFM phase at T$_O$. On raising H isothermally, de-arrest of group y will start at the field 2 indicated by horizontal arrow, and de-arrest will be complete on exiting the T${_K}$ band at horizontal arrow 4. Similarly, for FC along path 3, group x and y regions will transform to FM on cooling to T$_o$, and de-arrest of group z will start at 3. For FC along path 4, entire sample is in FM phase at T$_O$. Following each of the four FC paths depicted in Fig. 1(a) will thus result in M-H curves shown in the main panel of Fig. 1(b); de-arrest will start at the corresponding points during isothermal increase of field, and will be completed at the highest H end of the (H${_K}$, T${_K}$) band.

        On the contrary, if the two bands are correlated (i.e. the high-T end of the (H*, T*) band and the high-T end of the (H${_K}$, T${_K}$) band correspond to the same local region) then the frozen AFM regions will start getting de-arrested as H rises into the lower end of the (H${_K}$, T${_K}$) band independent of the field in which the sample is cooled. The field at which de-arrest starts depends only on the temperature at which H is raised isothermally. The de-arrest will be complete when all the frozen AFM regions are de-arrested. Since freezing is less at higher cooling field, de-arrest stops at lower H when the sample is cooled in higher field. The inset of Fig. 1(b) shows a qualitative schematic of how M-H curves would look under cooling in different fields, if the two landscapes are correlated. This is qualitatively very distinct from what is expected in the case of anti-correlated bands.
        
        We now consider measuring magnetization under one more path in H-T space. We field-cool the sample, in different fields, to T$_O$ and isothermally raise the field to that used for path 4 (H$_4$) shown in Fig. 1(a). From this field (H$_4$), we raise the temperature with field held constant at H$_4$. The arrested AFM regions will get de-arrested at different temperatures; that cooled by path 3 will convert to FM starting at the T marked by vertical arrow 3. Similarly for cooling by path 1 and 2. The de-arrest will start at lower temperatures for cooling in lower field, but full conversion to FM will occur at the same temperature in each case. The M(T) curves will then be as shown in the main panel of Fig. 1(c). It is trivial to argue that if the two landscapes are correlated, then de-arrest will start at the same T in each case, but full conversion to FM would occur at higher T for cooling in the lower field. The M(T) curves would qualitatively resemble the schematic in the inset of Fig. 1(c) which is very different from the case when the two landscapes are anti-correlated. 
        
        It is easy, given the schematic bands in Fig. 1(a) to visualize magnetization measurements under other H-T paths. It is asserted that the case of correlated landscape and that of anti-correlated landscapes predict qualitatively distinct behavior.
        
        We have so far been considering the case where the low (H, T) ground state is FM; we now consider the case where the low (H, T) ground state is AFM. This corresponds to the observations in doped CeFe$_2$ and in (NdSrSm)MnO$_3$. All arguments made earlier go through with appropriate changes, the supercooling spinoidal band (H*, T*) and the kinetic arrest band (H$_K$, T$_K$) are depicted in Fig. 2(a). The frozen FM fraction is larger at higher H. De-arrest occurs in isothermal case by reducing H (however it may not be complete even at H = 0). Fig. 2(b) shows the qualitative M-H expected if the two landscapes are anti-correlated, while the inset shows these for the case when they are correlated. De-arrest can also be caused by heating, M(T) should be measured after lowering the field isothermally. Fig. 2(c) is the counterpart of Fig. 1(c); again qualitatively different behavior distinguishes the case of correlated and anti-correlated bands.  
        
        To conclude, if a region with above average T* has above average T$_K$ , then the M-H and M-T curves obtained by following specified paths in the (H, T) space are qualitatively distinct than if that region has below average T$_K$ (i.e. higher T* region has lower T$_K$). It would be important to understand why disorder influences T* and T$_K$ in contrasting (or similar) manner.

{ }

\newpage
\begin{figure}
\caption{(color online) Schematics of H-T phase diagrams as well as the corresponding M-H and M-T curves after cooling in different fields for the case with FM ground state. (a) shows the H-T diagram with the anti-correlated bands for the case where the high-T state is AFM and the low-T equilibrium phase is FM.  Only 3 regions (viz. x, y, z) out of the N regions represented by the quasi-continuous (H*, T*) band are shown. The high-T side of the (H*, T*) band is shown by continuous line followed by dashed, dotted and dash-dotted lines. The corresponding 3-regions and the dividing lines in the (H$_K$, T$_K$) band appear in reverse order because of anti-correlation. For the first measurement protocol, after FC in different H to T$_O$ following paths 1-4 (1 being ZFC path) the fields for onset of de-arrest during the isothermal increase in H (at T$_O$) are shown by horizontal arrows for the respective H. Along the path-1 the sample is arrested completely in AFM phase while along path 4 the sample completely converts to FM phase during FC so does not have arrested phase. The variation in M resulting from this isothermal increase in H is sketched in main panel of (b). Steep increase in M starting at progressively higher H is due to de-arrest for the respective FC paths starting at higher H, as depicted by horizontal arrows in (a). The inset of (b) shows the expected M-H behavior for suitably sketched correlated bands (not shown). (c) depicts the next measurement protocol, after FC along paths 1-3 of (a) the H is isothermally raised to that of the path 4 (H$_4$), the M is measured while increasing the T from T$_O$. The resulting M-T is shown in the main panel. The abrupt change in the slope of M occurring at progressively higher T indicates the increase in the T at which de-arrest begins (during heating) at H$_4$, for the respective FC paths, as depicted by vertical arrows in (a). The inset of (c) shows the M-T curves for suitably sketched correlated bands (not shown).}
\label{fig1}
\end{figure}

\begin{figure}
\caption{(color online) (a), (b) and (c) are the counter parts of figures 1(a), 1(b) and 1(c) respectively, for the case when the low-T state is AFM. M-H curves of (b) are obtained by FC in different fields to T$_O$, and then reducing the the field to zero isothermally. M-T curves of (c) are obtained by reducing the field at T$_O$ to a finite value, and then heating in the same constant field. As seen from the insets of (b) and (c) correlated T$_K$ and T* bands would give qualitatively distinct M-H and M-T curves.}
\label{fig2}
\end{figure}
\begin{figure}
	\centering
		\includegraphics{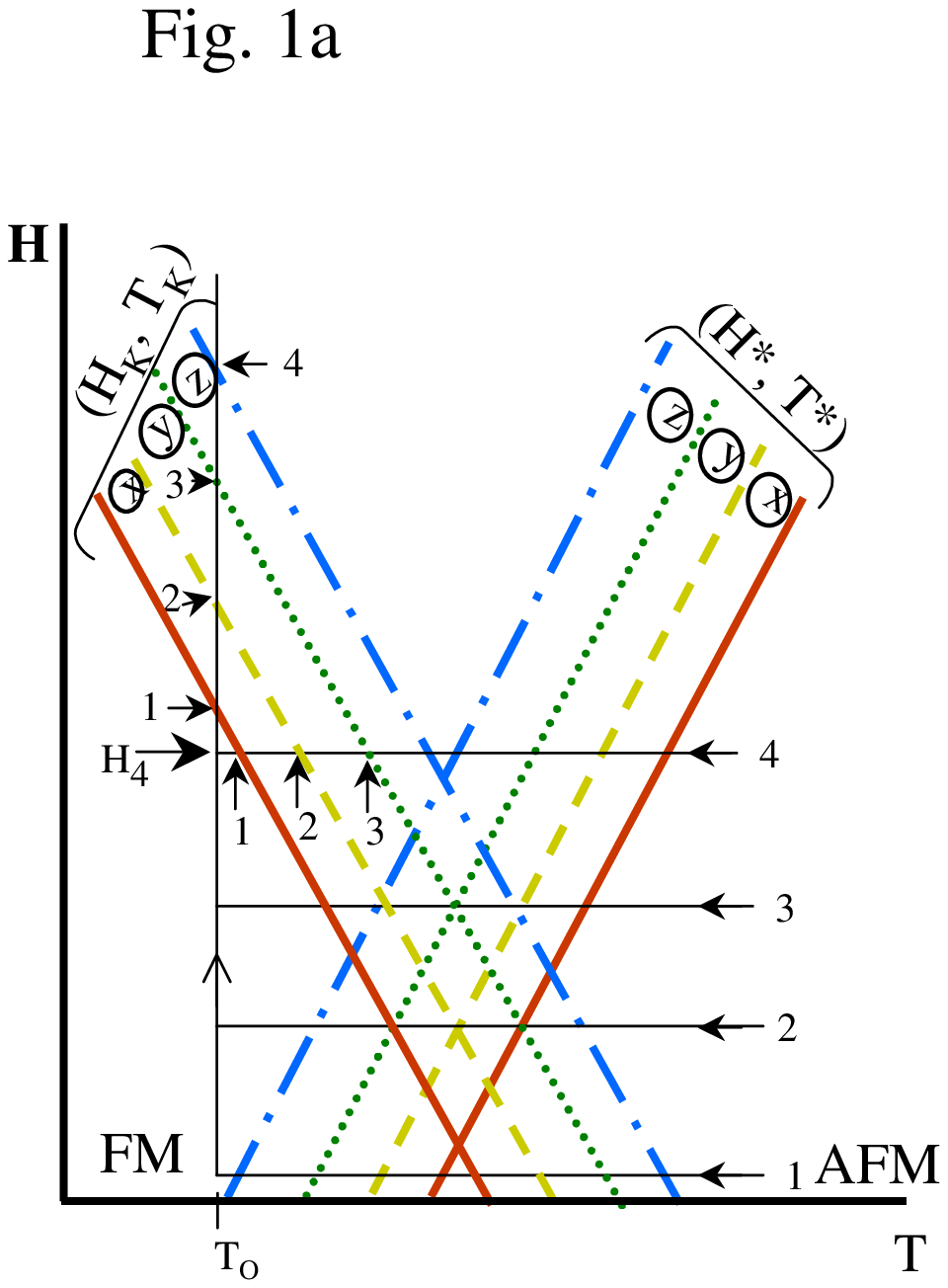}
	\label{fig:Fig1a}
\end{figure}
\begin{figure}
	\centering
		\includegraphics{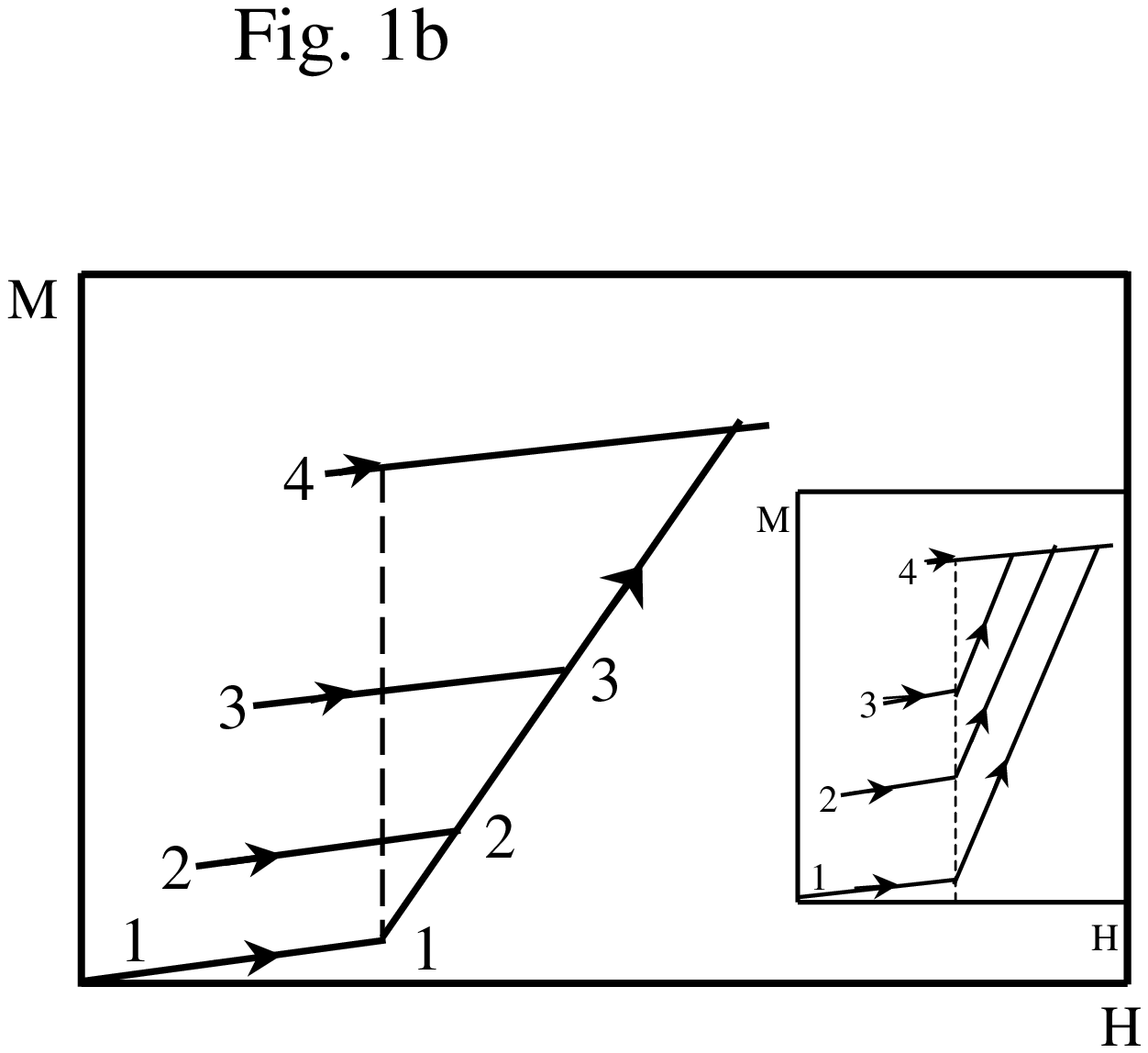}
	\label{fig:Fig1b}
\end{figure}
\begin{figure}
	\centering
		\includegraphics{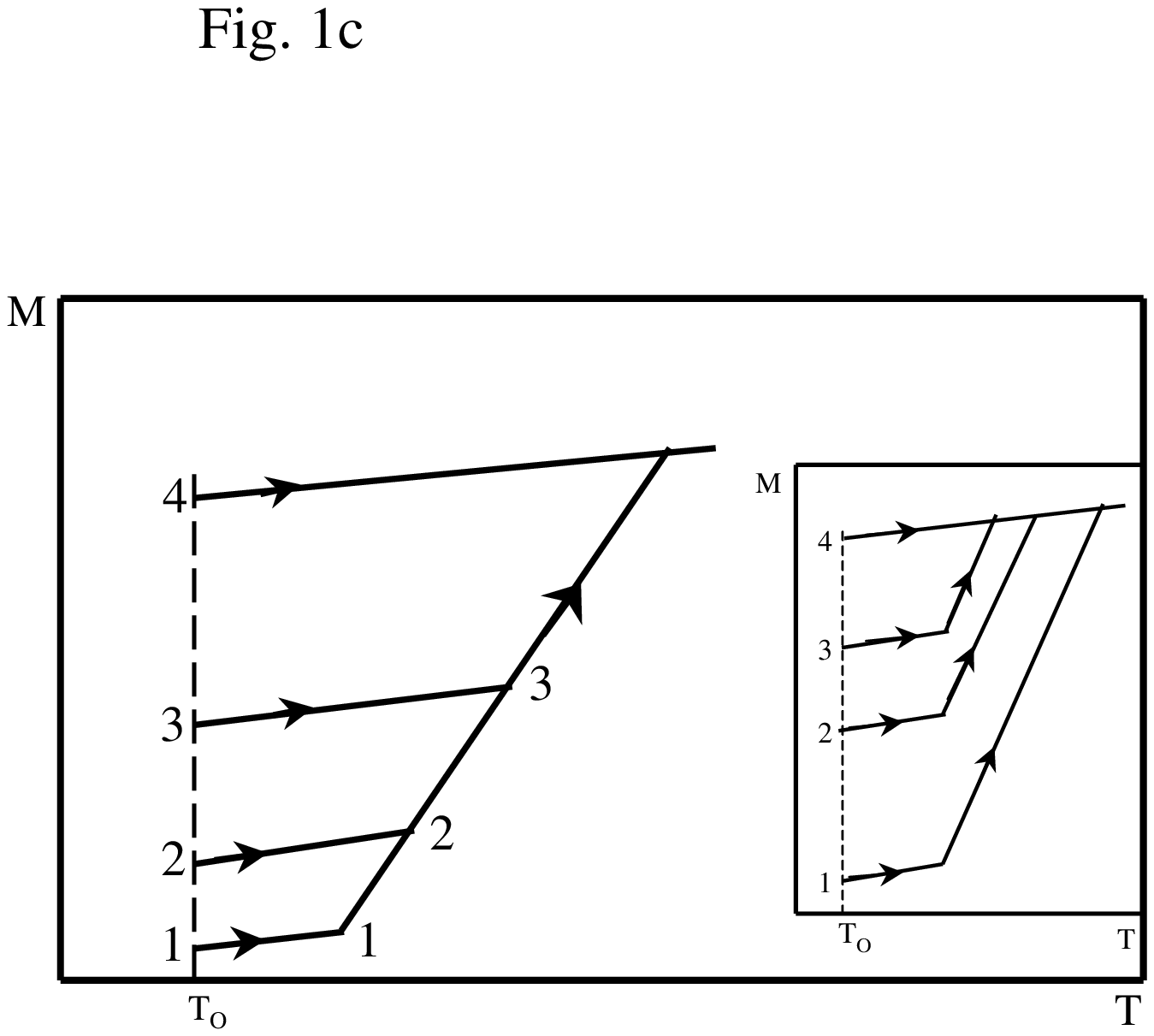}
	\label{fig:Fig1c}
\end{figure}
\begin{figure}
	\centering
		\includegraphics{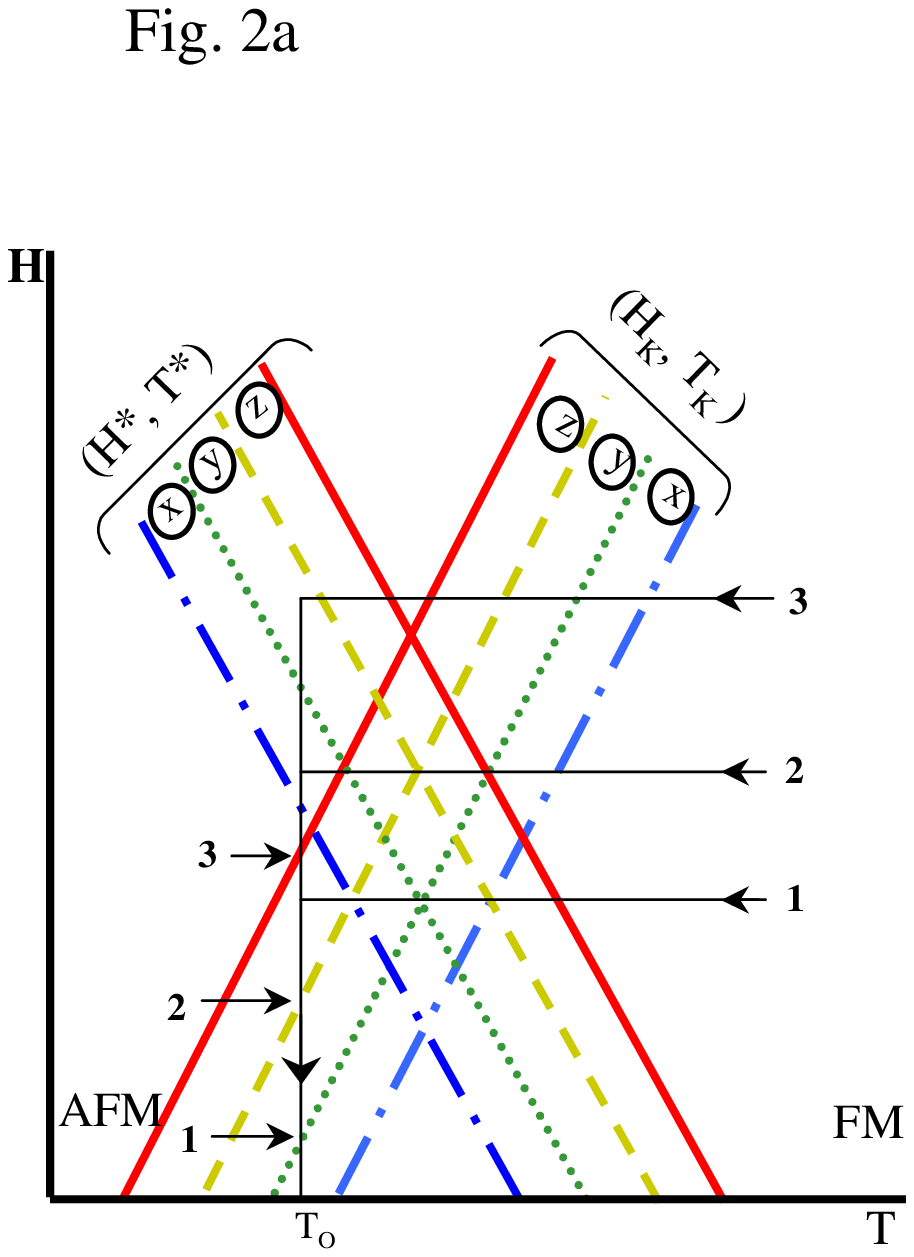}
	\label{fig:Fig2a}
\end{figure}
\begin{figure}
	\centering
		\includegraphics{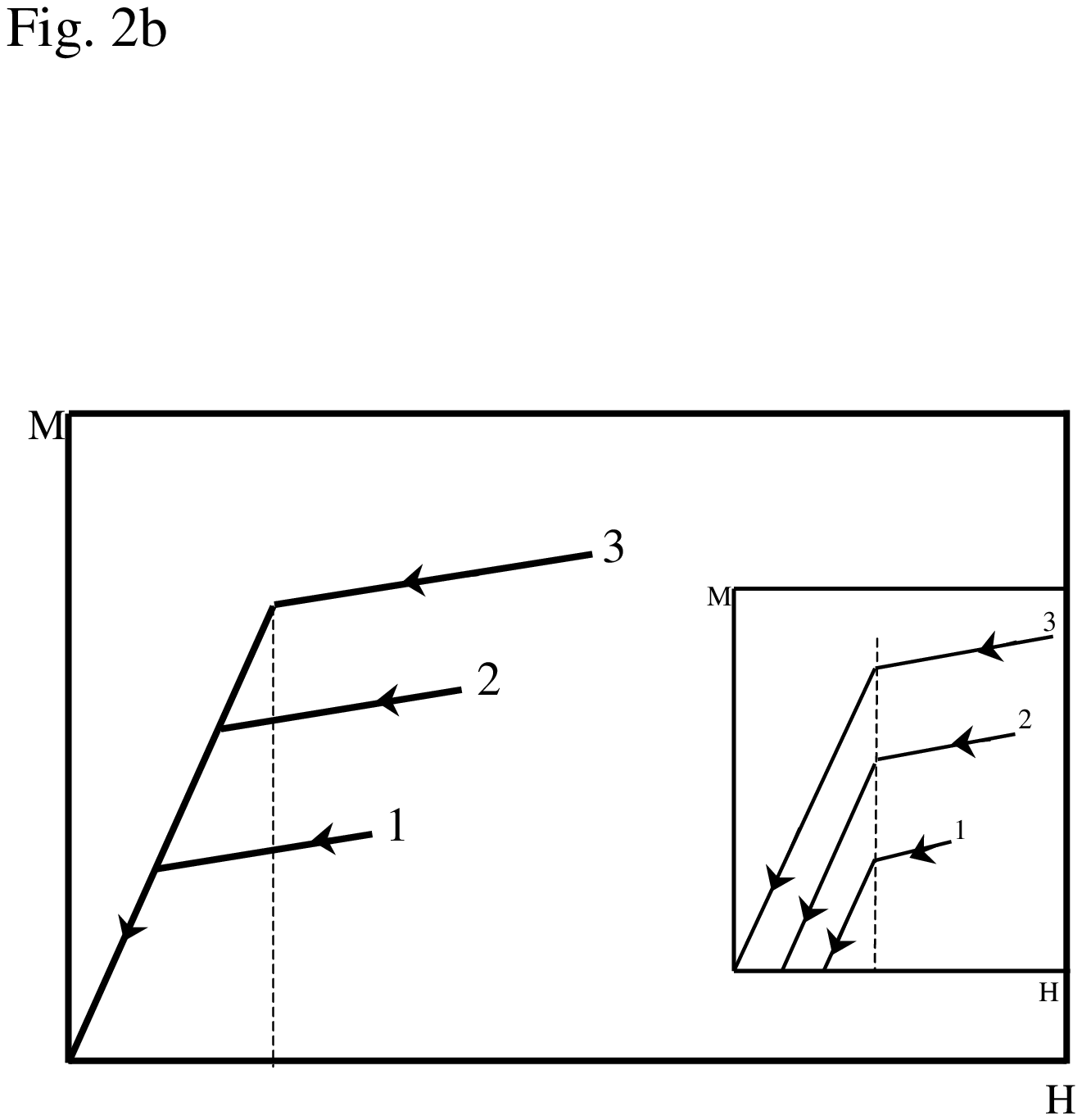}
	\label{fig:Fig2b}
\end{figure}
\begin{figure}
	\centering
		\includegraphics{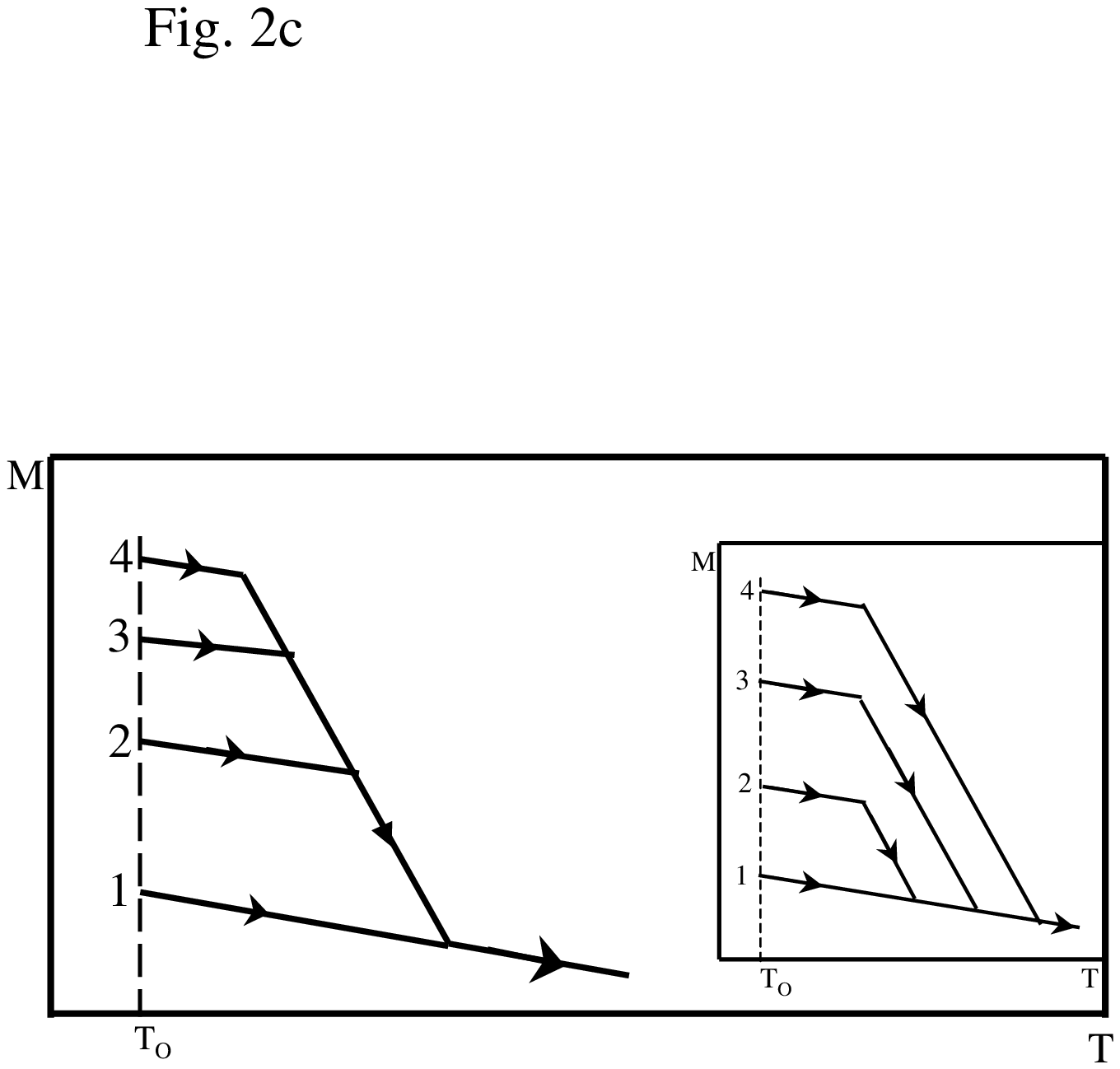}
	\label{fig:Fig2c}
\end{figure}


\begin{thebibliography}{ }
\bibitem[11]{ref1}The metastable kinetically arrested (or glass) state is different from the metastable supercooled state in that the latter (but not the former) will undergo a metastable to stable transformation on lowering of temperature.
\end{thebibliography}
\end{document}